# Quantifying the visual impact of wind farm lights on the nocturnal landscape

Salvador Bará [1] and Raul C. Lima [2,3, *]

[1] *Former profesor titular (retired) at Universidade de Santiago de Compostela (USC), Santiago de Compostela, 15782 Galicia (Spain, European Union)*

[2] *Física, Escola Superior de Saúde, Politécnico do Porto, Portugal*

[3] *IA – Instituto de Astrofísica e Ciências do Espaço, Univ Coimbra, Portugal*

e-mail: salva.bara@usc.gal, raulcpslima@ess.ipp.pt  (*) Corresponding author

**Abstract:** Wind farm lights are a conspicuous feature in the nocturnal landscape. Their presence is a source of light pollution for residents and the environment, severely disrupting in some places the aesthetic, cultural, and scientific values of the pristine starry skies. In this work we present a simple model for quantifying the visual impact of individual wind turbine lights, based on the comparison of their brightness with the brightness of well-known night sky objects. The model includes atmospheric and visual variables, and for typical parameters it shows that medium-intensity turbine lights can be brighter than Venus up to ∼4 km from the turbine, brighter than α CMa (the brightest star on the nighttime sky) until about ∼10 km, and reach the standard stellar visibility limit for the unaided eye ($m_v = +6.00$) at ∼38 km. These results suggest that the visual range of wind farms at nighttime may be significantly larger than at daytime, a factor that should be taken into account in environmental impact assessments.

## 1. Introduction

The need to reduce dependence on fossil fuels has fostered the development of renewable energy sources. This process has been accelerated in the last years due to the pressing urgency to address anthropogenic climate change and achieve higher levels of energy sovereignty. Among renewable sources, wind power energy is nowadays a crucial player.

The installation of new wind power facilities, both onshore and offshore, has not come without problems. Wind farms generate a wide range of environmental impacts [42,64], including but not limited to serious avian [30,34,50], and bat fatalities due to collisions [37,62,66,73] as well as changes in habitat use [41,67]. The sustainability of large offshore wind farms, planned or in construction, has been subjected to critical review in some recent European evaluations [20,45].

Besides their effects on biodiversity, wind farms also affect humans through the combined impacts of noise, lights (direct obstruction lights and stroboscopic effects of rotating blade shadows), and visual landscape degradation [48]. The annoyance produced by wind farm lights on neighboring communities has deserved growing attention in recent times [11,58,59,63].

The visual landscape degradation produced by wind farms has been evaluated mostly for daytime, based on turbine visibility estimates (limited by the contrast luminance thresholds in daylight) combined with different spatial aggregation metrics, see e.g. [31,36]. Comparatively less attention has been given to the deleterious effects of wind farm lights on the nighttime landscape (Fig. 1). The nightscape is an essential element of the human experience, whose cultural, social, scientific, and aesthetic values are assets of the intangible heritage of humankind [49]. As set forth by the Natural Sounds and Night Skies Division of the USA National Parks Service "a naturally dark night sky is more than a scenic canvas; it is part of a complex ecosystem that supports both natural and cultural resources" [56]. Borrowing from Rich and Longcore [61] on conservation planning, it can certainly be said that daytime landscapes are "only half the story–the daytime story".

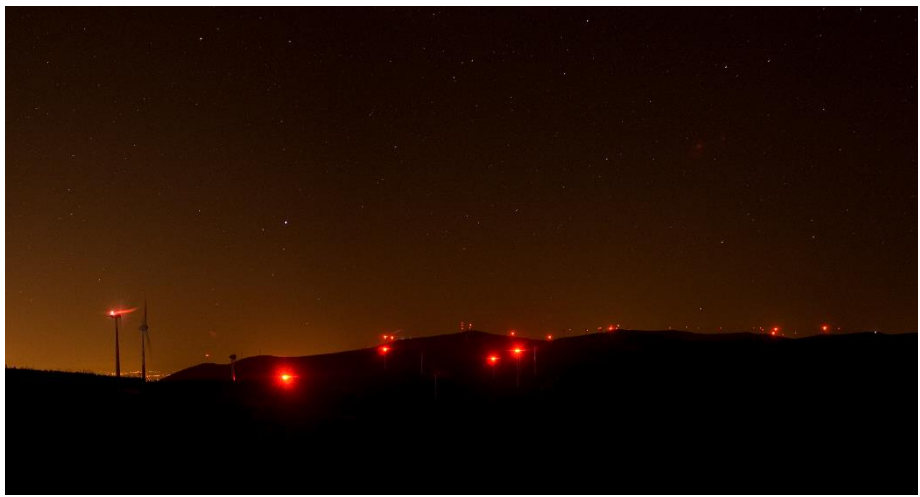

**Fig. 1.** Nigthtime landscape with wind farm obstruction lights in Miranda do Corvo, Serra da Lousã, Portugal (40°02'42.98"N, 8°16'30.84" W). Image credit: Raul C. Lima.

In this work we present a model for quantifying the primary visual effect of wind farm lights on the nocturnal landscape. It is based on considering wind farm lights as artificial stars and applying to them the metrics used in visual astronomy to quantify their perceived brightness. That way they can be compared with the stars and other natural bodies present on the sky, facilitating an easy and intuitive evaluation of the disruption caused to the pristine nightscape. The model incorporates atmospheric and perceptual parameters. Although it is formulated in terms of human-based photometric (in lighting engineering language) or visual band (in astronomical language) quantities, its generalization for arbitrary spectral distributions and observation bands is immediate. Wildlife shows a rich variety of spectral response curves [46], and several species have been reported to be able to use celestial cues, including individual stars, for orientation during the night (for a review, see [22]). The visibility model here developed provides the basic building block from which overall visual impact assessments can be derived with the help of different spatial aggregation metrics.

## 2. Methods

2.1. Retinal images of resolved and unresolved objects

The visual brightness of an object depends on several radiometric and anatomical factors, besides the physiological and neural ones. The basic inputs to the visual system are the photon catches of the retinal photoreceptor cells. The number of photons captured by an individual photoreceptor in each wavelength interval $[\lambda, \lambda + \Delta\lambda]$ during the time $\Delta t$ is proportional to $S(\lambda)\, E'(\lambda)\, \Delta\lambda\, \Delta t$, where $S(\lambda)$ is the spectral sensitivity of the photoreceptor and $E'(\lambda)$ is the spectral irradiance of the object's retinal image at the photoreceptor location. Wavelength values are here referred to vacuum. Hereafter we explicitly make reference to human eyes. However, the basic equations described below can also be applied to "camera-like" eyes of other species, many examples of which exist in nature (for a review, see [38]), by adapting the corresponding geometrical, optical, and spectral sensitivity parameters.

The spectral irradiance is defined as the radiant flux (photons·s$^{-1}$) per unit surface and unit wavelength interval, and is measured in photons·s$^{-1}$·m$^{-2}$·nm$^{-1}$. The spectral irradiance $E'(\lambda)$ at each retinal location can be approximately described by:

$$E'(\lambda) = T_\mathrm{e}(\lambda)\, L(\lambda)\, \Omega\, \frac{A_\mathrm{p}}{A_\mathrm{i}} \cos\theta \qquad (1)$$

where $T_\mathrm{e}(\lambda)$ is the spectral transmittance of the ocular media (unitless), $L(\lambda)$ is the object radiance (photons·s$^{-1}$·m$^{-2}$·sr$^{-1}$·nm$^{-1}$), $\Omega$ is the solid angle (sr) subtended by the object as seen from the observer, $A_\mathrm{p}$ is the area of the input pupil of the eye (m$^2$), $A_\mathrm{i}$ is the area of the retinal image (m$^2$), and $\theta$ is the angle between the direction in which the object lies and the line perpendicular to the eye pupil. Eq. (1) can be equivalently rewritten in terms of $E(\lambda) = L(\lambda)\, \Omega \cos\theta$, the spectral irradiance produced by the object on the eye pupil, as:

$$E'(\lambda) = T_\mathrm{e}(\lambda)\, E(\lambda)\, \frac{A_\mathrm{p}}{A_\mathrm{i}} \qquad (2)$$

In a perfect imaging system, according to geometrical optics, the image would be an exact scaled replica of the object. The area of the image would be proportional to the solid angle subtended by the object, $\Omega$, such that $A_\mathrm{i} = \kappa\, \Omega$, being $\kappa$ a constant (units m$^2$) independent from the object. For a perfect imaging system, then, the object feature that determines the input to the photoreceptor cells is its spectral radiance, $L(\lambda)$, since the solid angle $\Omega$ cancels out in Eq. (1):

$$E'(\lambda) = T_e(\lambda)\, L(\lambda)\, \frac{A_p}{\kappa} \cos\theta \tag{3}$$

and the remaining factors only depend on the eye.

Human eyes, however, are not perfect optical instruments. The eye optics deforms the ideal images to a bigger or lesser extent due to diffraction by the finite size of the eye pupil [13], refractive errors including both classical ametropies and high-order optical defects [15,28,43,44,47,53,54,55,60,69,70], and intraocular scattering due to small-scale inhomogeneities of the eye media [8,71,72]. The retinal image in a real eye is no longer an exact scaled replica of the object itself.

The actual image of an object on the human retina is given by the two-dimensional convolution of the ideal geometric image with the point-spread function of the eye, PSF [25]. The PSF is the function that describes how the eye optics deforms the image of an ideal point source. Human eyes present a high variability of PSF sizes and shapes. This convolution gives rise to a new irradiance distribution that combines the features of the convolved functions. The retinal image can be interpreted as a 'blurred' version of the perfect geometrical image, in which each point has been replaced by a PSF proportional to the object radiance at that point and the resulting irradiances have been added up.

In the limiting case of *well resolved objects*, that is, when the angular size of the object is much larger than the angular size of the PSF, the result of the convolution is a slightly blurred version of the ideal geometrical image, and Eq. (3) still approximately applies.

Conversely, when the angular size of the object viewed by the observer is substantially smaller than the angular size of the eye's PSF, as it happens e.g. with a star or distant streetlight, the retinal image is essentially equal to the PSF itself. In such cases, one may speak of *unresolved objects*. The retinal images of unresolved objects located in different directions of the central visual field are just replicas of the PSF centered in different retinal points. All these images have the same shape and size, only differing in brightness [4,54,55].

This result has important visual consequences. One of them is that the main physical factor determining the perceived brightness of an unresolved object is not the intrinsic object's radiance, as in the well-resolved case, but the irradiance $E(\lambda)$ it produces on the eye pupil. This happens because the area of the retinal image is no longer proportional to the object's solid angle, as in case of a perfect system ($A_i = \kappa\, \Omega$), but it is constant and equal to the area of the PSF, $A_i = A_{\text{PSF}}$. For an unresolved object Eq. (2) becomes:

$$E'(\lambda) = T_e(\lambda)\, E(\lambda)\, \frac{A_p}{A_{\text{PSF}}} \tag{4}$$

in which the object intervenes through $E(\lambda)$, since the remaining factors depend only on the eye.

The pupil size depends, among other factors, on the ambient light level and the age of the observer, with large variability between individuals [40]. The PSF size, which also shows large inter-individual variability in healthy eyes due to the differences in uncompensated ametropies, high-order eye refractive defects, and intraocular scattering commented above, depends on a non-monotonic way on the pupil size. Diffraction makes the PSF size to increase when the pupil size decreases, being dominant for pupils smaller than ~2 mm diameter; the contribution of the residual uncompensated ametropies, higher-order defects and intraocular scattering, in turn, varies in the opposite way, typically determining the PSF size for pupils of diameter ~4mm and larger. Consequently, the best optical quality in healthy human eyes is usually achieved with pupils in the range of 2-3 mm diameter. For graphical depictions of actual PSFs see [5,54,55].

In terms of human visual (photometric) quantities, the irradiances and radiances outside the eye correspond to illuminances and luminances, respectively. The illuminance $E_v$, measured in lux (lx), is the integral over wavelengths of the spectral irradiance $E(\lambda)$ weighted by the photopic luminous efficiency function $V(\lambda)$ and multiplied by the luminous efficacy constant 683 lm/W [16]. The luminance $L_v$, measured in candela per square meter (cd·m$^{-2}$), is the analogous integral applied to the spectral radiance $L(\lambda)$. The cd is the unit of luminous intensity, $I_v$ (1 cd=1 lm·sr$^{-1}$), being the only basic unit of the International System [10] whose definition is strictly tied to human visual perception. The cd links perceptual to physical stimuli, and by its own definition it takes implicitly into account the transmittance of the ocular media, $T_e(\lambda)$, the spectral sensitivity of the photoreceptors, $S(\lambda)$, and the basic neural processes related to the perception of luminance.

2.2. Illuminance produced by wind turbine lights on the eye pupil of the observer

The illuminance $E_v$ produced by an unresolved wind turbine light on the eye pupil of an observer located a distance $r$ away is:

$$E_v(r,\theta) = T_0(r)\, I_v(\alpha)\, \frac{\cos\theta}{r^2} \qquad (5)$$

where $T_0(r)$ is the transmittance of the atmosphere between the wind farm light and the eye, $I_v(\alpha)$ is the luminous intensity of the wind turbine light sent towards the observer, being $\alpha$ the direction in which the observer is located as seen from the turbine, and $\theta$ is the angle between the normal to the pupil and the direction where the wind turbine light is located as seen from the observer. This equation, published in simplified form by Allard one and a half century ago [2,14] stems from the basic definition of these photometric quantities and the properties of light propagation in attenuating media. The formulation of Eq. (5) in terms of the luminous intensity of the wind turbine lights $I_v$ is particularly useful, because this is the photometric quantity specified for the different types of obstruction lights in the ICAO recommendations [32], and whose values have been generally adopted by the wind farm legislations, e.g. AESA [1].

For light propagation paths at angles close to the horizontal, and direct visual fixation on the turbine light ($\theta = 0°, \cos\theta = 1$), Eq. (5) becomes

$$E_v(r) = I_v\, \frac{e^{-kr}}{r^2} \qquad (6)$$

where $I_v = I_v(\alpha)$, and the atmospheric transmittance $T_0(r)$ is described by the Bouger-Lambert exponential law $T_0(r) = e^{-kr}$, being $k$ the combined molecular and aerosol attenuation coefficient per unit length (m$^{-1}$) at the average altitude above sea level of observer and lights.

2.3. How bright are wind turbine lights compared to the stars and planets of the natural sky?

The visual brightness of the stars and other unresolved point-like sources on the sky is commonly reported in astronomy in terms of "astronomical magnitudes". The astronomical magnitude is a negative logarithmic scale for expressing the in-band irradiance produced by a star, relative to some reference irradiance. This scale was qualitatively introduced by Hipparchus (c. 190–c. 120 BC) and formalized in the 19th century [57]. The magnitude scale is also applied to illuminances $E_v$, taking as traditional reference the illuminance produced by the star Vega (α Lyr) at the top of the terrestrial atmosphere, $E_{v,\text{ref}} = E_{v,\text{Vega}} = 2.54 \times 10^{-6}$ lx [3,6,7]. According to the basic definition of this scale, an unresolved celestial object producing an illuminance $E_v$ (in lx) at the top of the terrestrial atmosphere has a visual magnitude $m_v$ given by

$$m_v = -2.5 \log_{10}\left(\frac{E_v}{E_{v,\text{ref}}}\right) \tag{7}$$

Conversely, the illuminance at the top of the atmosphere in terms of the magnitude is

$$E_v = E_{v,\text{ref}} \times 10^{-0.4\, m_v} \quad \text{(lx)} \tag{8}$$

It is conventionally but somehow arbitrarily accepted that an average observer may detect stars up to $m_v \approx 6.0$. As a matter of fact, the limiting visual magnitude of the unaided eye depends on many factors, including the luminance contrast threshold of the observers at the luminance adaptation level they are experiencing, the state of the atmosphere, and the artificial skyglow and glare (two effects of light pollution) at the observer location. For a detailed analysis see [8,17,65].

When a star is observed from the ground, its extra-atmospheric illuminance is reduced due to the attenuation undergone by the light rays along their path through the whole atmosphere. This can be accounted for by an atmospheric transmittance term

$$T_{\text{atm}}(z) = \exp\{-M(z)\,\tau\} \tag{9}$$

where $z$ is the angle from the star to the zenith, $\tau$ is the atmospheric vertical optical depth, and $M(z)$ is the air mass number. The atmospheric optical depth is given by $\tau = \tau_m + \tau_a$, where $\tau_m$ and $\tau_a$ are the molecular (MOD) and aerosol (AOD) optical depths, respectively, defined in terms of the corresponding vertical profiles of the molecular $k_m(h)$, and aerosol, $k_a(h)$, extinction coefficients [35] as:

$$\tau_i = \int_{h=0}^{\infty} k_i(h)\, dh \qquad i \in \{m, a\} \tag{10}$$

For an exponential atmosphere in which $k_\mathrm{m}$ and $k_\mathrm{a}$ decrease exponentially with the altitude $h$, with scale heights $H_\mathrm{m}$ and $H_\mathrm{a}$, respectively, we have

$$\tau_i = \int_{h=0}^{\infty} k_i(0)\, e^{-h/H_i}\, \mathrm{d}h = k_i(0)\, H_i \qquad i \in \{\mathrm{m}, \mathrm{a}\} \tag{11}$$

being $k_i(0)$ the value of the attenuation coefficients at ground level. Typical values for the exponential scale heights are $H_\mathrm{m} = 8$ km and $H_\mathrm{a} = 1.5$ km. The value of $\tau_\mathrm{m}$ at sea level at the center of the visible spectrum is about 0.09-0.11, see [68] for detailed expressions, and the aerosol optical depth $\tau_\mathrm{a}$ may range typically from 0.1 or smaller for clear atmospheres to 0.5 and larger for more turbid ones, being able to reach much higher values [19,24,29]. Note that the coefficient $k$ appearing in Eq. (6) can be written as

$$k = k_\mathrm{m}(0) + k_\mathrm{a}(0) = \frac{\tau_\mathrm{m}}{H_\mathrm{m}} + \frac{\tau_\mathrm{a}}{H_\mathrm{a}} \tag{12}$$

Regarding the airmass factor, for zenith angles not extremely close to the horizon its value can be calculated as $M(z) = 1/\cos z$. The number of air masses increases very quickly at angles close to the horizon, for which more accurate expressions shall be used [33]. For the zenith, $M(0°) = 1$.

The atmospheric transmittance in Eq. (9) can be expressed as an equivalent extinction value $m_\mathrm{ext}(z)$ in magnitudes. The magnitude $m_\mathrm{v}(z)$ of a star observed from ground at a zenith angle $z$ (angle above the horizon $90° - z$) is:

$$m_\mathrm{v}(z) = -2.5 \log_{10}\left[\frac{T_\mathrm{atm}(z)\, E_\mathrm{v}}{E_\mathrm{v,ref}}\right] = -2.5 \log_{10}\left(\frac{E_\mathrm{v}}{E_\mathrm{v,ref}}\right) - 2.5 \log_{10}[T_\mathrm{atm}(z)] \tag{13}$$

that can be rewritten as

$$m_\mathrm{v}(z) = m_\mathrm{v} + m_\mathrm{ext}(z) \tag{14}$$

being $m_\mathrm{v}$ the extra-atmospheric magnitude given by Eq. (7), and $m_\mathrm{ext}(z)$ the extinction term $m_\mathrm{ext}(z) = 2.5\, M(z)\, \tau \log_{10}(e)$. Recall that larger (= more positive) values of $m_\mathrm{v}(z)$ correspond to dimmer objects, due to the negative sign of the log scale magnitude definition.

The brightness of the wind farm lights can also be expressed in astronomical magnitudes, allowing that way comparing them with the natural stars. This could be done in a naïve way by directly applying Eq. (7) to the illuminance produced by the lights on the observer's eye pupil, $E_\mathrm{v}(r)$, given in Eq. (6). There is, however, an issue that shall be kept in mind. Whereas the astronomical magnitudes refer to the irradiance produced by a celestial object *at the top* of the atmosphere, the light from wind farms reaches the observer after propagating some finite distance nearly horizontally *at the bottom* of the atmosphere. Furthermore, the brightness of a star seen from ground is not constant, but depends on its altitude above the horizon, or, equivalently, on its corresponding zenith angle $z$ as set forth in Eq. (13). Comparing the visual appearance of wind farm lights with the appearance of stars requires choosing first a reference altitude above the horizon at which the comparison stars are seen.

Once the reference zenith distance $z$ is chosen, one can easily assign to the wind farm light the extra-atmospheric magnitude $m_\mathrm{v}$ of a star whose brightness at this $z$ would be the same as the brightness of the wind farm light perceived by the observer. According to Eqs. (6) and (7) this magnitude is:

$$m_\text{v} = -2.5 \log_{10} \left[ \frac{I_\text{v}\, e^{-kr}}{r^2\, T_\text{atm}(z)\, E_\text{v,ref}} \right] \quad (15)$$

Note that the transmittance $T_\text{atm}(z)$ is in the denominator, since we are calculating the extra-atmospheric irradiance that would result in the irradiance $E_\text{v}(r)$ at ground level after propagating through the atmosphere at a zenith angle $z$. This can alternatively be interpreted as using a reduced $T_\text{atm}(z)\, E_\text{v,ref}$ reference illuminance for establishing the 'zero point' of a new magnitude scale defined on irradiances at ground level, not at the top of the atmosphere.

Regarding the choice of $z$, comparing the wind farm lights with stars at the zenith (altitude 90°, $z=0°$, $M(0°) = 1$) is always an option, although to perform in practice this comparison the observers should successively direct their gaze horizontally to the wind farm and vertically to the zenith sky, because the possibility of simultaneous viewing (although theoretically possible, given the size of the monocular field of view of the human eye along the vertical axis, ~60° upward and ~75° downward) would require that both light sources were imaged in diametrally opposed locations of the peripheral retina. Comparing wind farm lights with stars at the horizon would neither be a practical choice, since the attenuation of starlight in that direction is usually extremely high excepting for very clear atmospheres ($\tau \ll 0.05$), due to the large number of air masses, $M(90°) \approx 38$ [33]. For instance, for $\tau = 0.2$ the extinction at the horizon is of order $m_\text{ext}(90°) = +8.25$ magnitudes. For our present purposes an intermediate altitude above the horizon, well within the visual field of an observer looking horizontally at the lights, is appropriate. For the following sections we will use a reference altitude of 30° ($z = 60°$, $M(60°) = 2.0$). The conversion of our results for other possible choices of the altitude above the horizon of the comparison stars is immediate.

## 3. Results

Figure 2 shows the equivalent top-of-the-atmosphere (TOA) astronomical magnitude $m_\text{v}$ of a wind farm light (for a reference altitude of 30° degrees above horizon, $z = 60°$, $M(60°) = 2.0$), as a function of the distance to the viewer, $r$. The atmospheric parameters are $\tau_\text{m} = 0.10$, $\tau_\text{a} = 0.20$, $H_\text{m} = 8000$ m and $H_\text{a}=1500$ m. The results were calculated for nighttime obstruction lights of medium-intensity, $I_\text{v} = 2000$ cd, and for two levels of low-intensity, $I_\text{v} = 200$, and $I_\text{v} =40$ cd [1,32]. The extinction coefficient at ground level is $k = 1.46 \times 10^{-4}$ m$^{-1}$, corresponding to a daytime visual range of ~26 km. The horizontal lines show the astronomical magnitudes in the Johnson-Cousins V band [9] of several conspicuous objects on the sky, namely the Moon, $m_\text{v} = -12.73$ (full Moon at mean distance from Earth, near opposition but not including the opposition surge, [3] p. 144), Venus, $m_\text{v} = -4.22$ (mean magnitude of Venus at maximum elongation, [3] p. 144), and the star Sirius (α CMa), $m_\text{v} = -1.45$ ([3] p. 240). The standard human star visibility limit with the unaided eye, $m_\text{v} = +6.00$, is also shown. In the Johnson-Cousins V band the magnitude of the star Vega (α Lyr) is usually set at $m_v = +0.03$. Fig. 2 (a) shows the magnitude values within the 0.01-50 km distance range, whereas Fig. 2 (b) shows an enlarged view of the first 5 km from the lights.

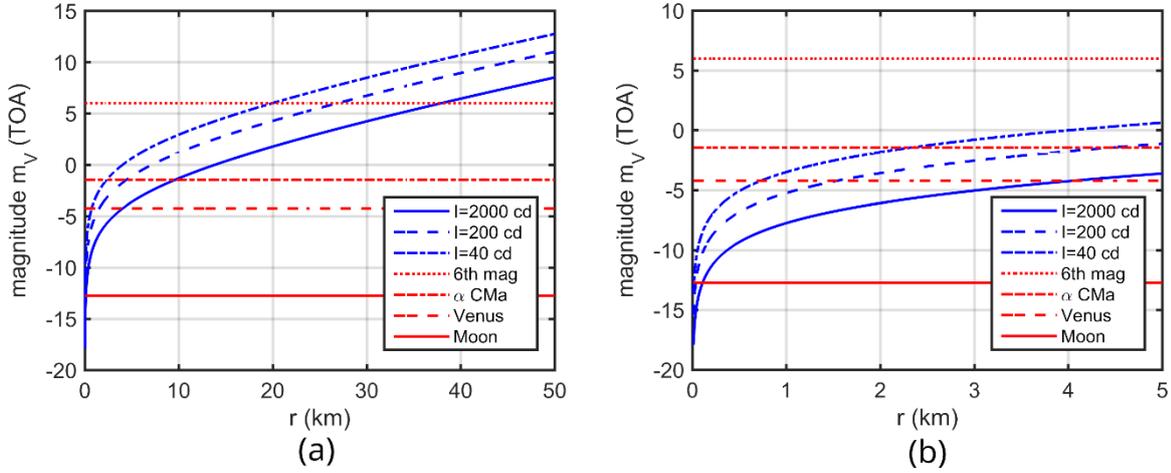

**Fig. 2.** Equivalent top-of-the-atmosphere (TOA) astronomical magnitude $m_v$ of wind turbine lights seen from distances $r$, Eq. (15), for lamps of luminous intensities $I_v = 2000$, 200, and 40 cd, and an atmosphere with AOD $\tau_a = 0.2$ (see text for details). (a) range 0.01-50 km, (b) enlarged view of the range 0.01-5.0 km.

It can be seen in Fig. 2 that the nighttime lights of $I_v = 2000$ cd widely used in wind turbines of height between 100 and 150 m [1] can be brighter than Venus up to 4 km from the turbine, brighter than α CMa up to about 10 km, and reach the visibility limit $m_v = +6.00$ at 38 km. These results suggest that the visual range of wind farms at nighttime in pristine sites (limited by the luminous intensity of the lamps and the atmospheric attenuation) may be significantly higher than at daytime (limited by the luminance contrast thresholds applied to sunlight scattered into the line of sight [12]).

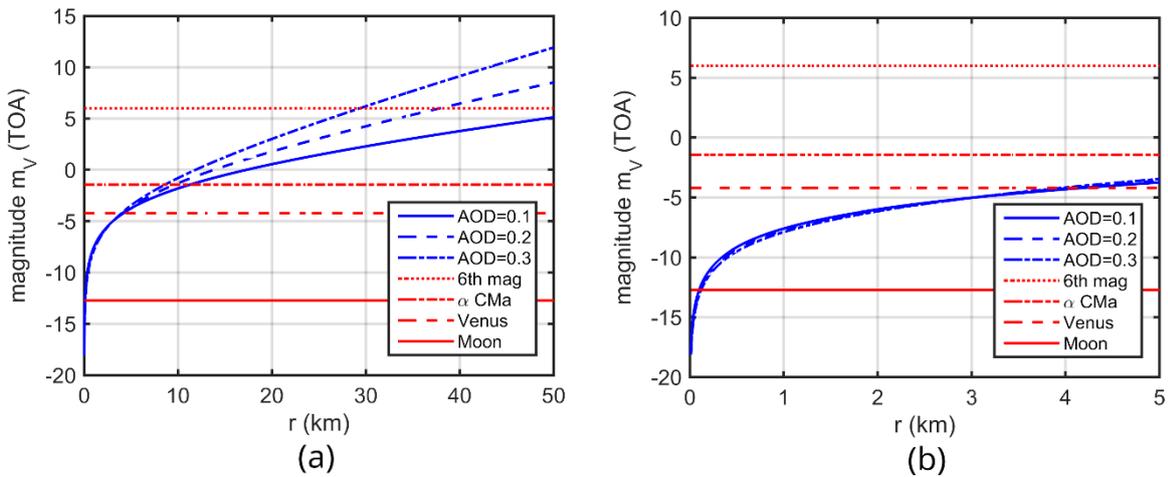

**Fig. 3.** Equivalent top-of-the-atmosphere (TOA) astronomical magnitude $m_v$ of wind turbine lights seen from distances $r$, Eq. (15), for lamps of luminous intensity $I_v=2000$ cd and atmospheres with AODs $\tau_a$= 0.1, 0.2, and 0.3 (see text for details). (a) range 0.01-50 km, (b) enlarged view of the range 0.01-5.0 km.

Figure 3 shows the magnitudes $m_v$ for lights of constant luminous intensity ($I_v$ = 2000 cd) and three different aerosol optical depths, $\tau_a$= 0.1, 0.2 and 0.3, corresponding to daytime visual ranges of 49 km, 26 km, and 18 km, respectively, the remaining parameters being the same as in Fig. 1. During the first kilometers from the lamps the change in visual magnitude is dominated by the geometrical factor $1/r^2$ in Eq. (15), with little influence of the horizontal-path atmospheric attenuation $e^{-kr}$, so the $m_v$ curves for different AOD are very close to each other. As the distance increases, the horizontal atmospheric attenuation becomes dominant and the values of $m_v$ increase almost linearly with $r$. The overall behavior is also function of the value of $T_{atm}(z=60°)$, the atmospheric attenuation for an equivalent star seen at 30° above the horizon (here, two air masses), the reason behind the fact that the curves do not fully overlap in the first few km from the sources.

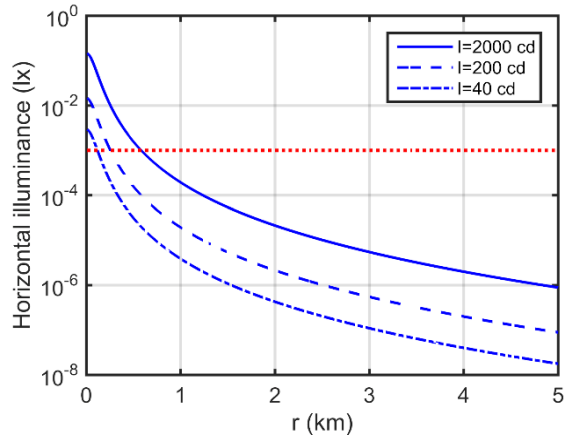

**Fig. 4.** Horizontal ground illuminance (in lx) produced by a single light of intensity 2000 cd located on a nacelle at $h_n$ =115 m above ground, with isotropic angular emission and under the same atmospheric conditions as in Fig. 2. The horizontal, dotted line corresponds to the illuminance produced by a typical moonless starry sky in conditions of astronomical night, ~0.001 lx. The variable $r$ in this figure corresponds to the horizontal distance from the base of the turbine, such that the total distance from the lamp to the ground observation point is $\left(r^2 + h_n{}^2\right)^{1/2}$.

Figure 4 shows the horizontal ground illuminance (in lx) produced by a single light of intensity 2000 cd located on a nacelle at 115 m above ground, with isotropic angular emission, and under the same atmospheric conditions as in Fig. 2. This illuminance is calculated using Eq. (5), by reinterpreting the angle $\theta$ as the zenith angle of the light source seen from the observer location. The constant, dotted line corresponds to the illuminance produced by a typical moonless starry sky in conditions of astronomical night, ~0.001 lx [51,52]. It can be seen that the horizontal ground illuminance produced by this mid-intensity light surpasses that of the moonless starry sky within the first 580 m from the base of the turbine, becoming two orders of magnitude smaller at a distance of 2.5 km. Recall that, unlike the retinal illuminances of non-overlapped lights perceived as individual objects, the ground illuminance produced by several lights is accumulative, hence the total ground illuminance at any point in the neighborhood of a wind farm shall be calculated as the sum of the contributions of all lamps. The direct luminous intensity of a single lamp, however, elicits significant visual responses up to much longer distances (see Fig. 2 and 3, and Section 4).

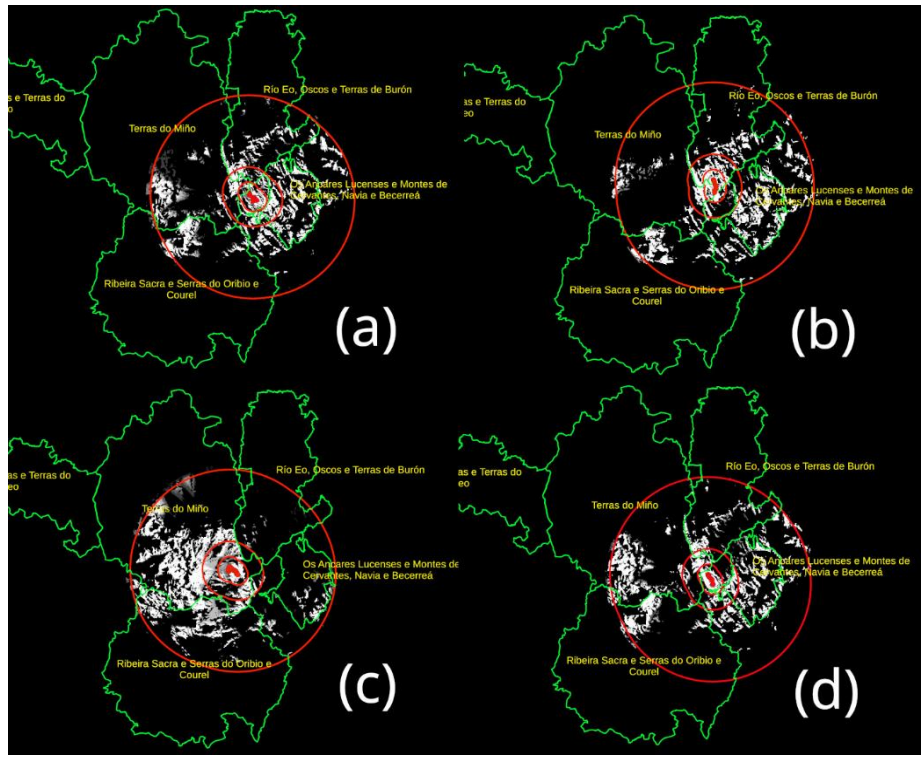

**Fig. 5.** Visual territorial impact of four windfarm projects in Galicia (kingdom of Spain, European Union). Project names, number of wind turbines ($N$) and nacelle heights ($h_n$): (a) Chao do Marco, $N$=8, $h_n$ =115 m; (b) Monteiro, $N$=8, $h_n$ =107 m; (c) Reboiro, $N$=11, $h_n$ =121 m; (d) Serra do Furco, $N$=7, $h_n$ =115 m. Graylevels indicate the number of wind turbines of each windfarm that can be seen from each pixel (from 0 to $N$). The red lines around each windfarm show the 4 km, 10 km, and 38 km range areas. The green borders show the limits of the surrounding Biosphere Reserves (clockwise, from the upper left, "Mariñas Coruñesas e Terras do Mandeo", "Terras do Miño", "Río Eo, Oscos e Terras de Burón", "Os Ancares Lucenses e Montes de Cervantes, Navia e Becerreá", and "Ribeira Sacra e Serras do Oribio e Courel". Shp layer copyright: management boards of the Reserves). Digital elevation model PNOA_MDT200_ETRS89_HU29 (Galicia) (PNOA 2010-2013 CC-BY scne.es). Maps elaborated with QGIS ver. 3.22.10-Białowieża (https://qgis.org/)

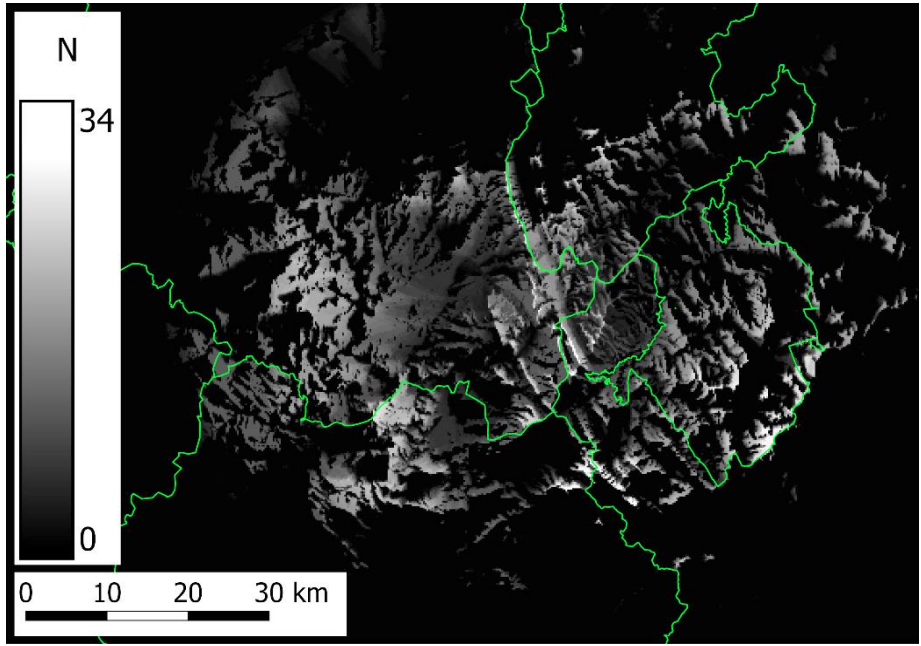

**Fig. 6.** Aggregated visibility map of the four wind-farm projects displayed in Fig. 5. The grayscale indicates the total number *N* of wind turbine nacelles seen from each location.

The fraction of territory within the visual range from which the windfarm lights can be seen depends on the local topography and on the height of the lights above ground level. While in flat areas the direct line of vision may be mostly unblocked across wide spans of territory, in mountainous regions it is generally limited by shadowing effects. However, since windfarms tend to be located on the highest elevations of mountain ridges, this fraction is generally very large. An example of the latter is shown in Fig. 5, where the visibility maps of four windfarm projects in the Eastern Mountains of Galicia (historic nationality and autonomous community in the kingdom of Spain, European Union) are displayed as graylevel rasters with the number of wind turbines from each windfarm that can be seen from each pixel. The red lines around each windfarm show the 4 km, 10 km, and 38 km range areas. These four windfarms are located relatively close to each other. Each one has particular visibility impacts, but the affected fraction of territory is in all cases remarkable. Their aggregated effect is shown in Fig. 6, at a spatial scale that allows discerning the variable but widespread visual impacts across these Biosphere Reserves region.

## 4. Discussion

The results obtained using the main equation of this work, Eq. (15), show that wind turbine lights can compete with natural sky objects up to distances in the range of tens of km. This has a non-negligible effect on the nocturnal landscape, particularly so because these lights are normally seen close to the horizon, towards which very often the human direction of gaze is oriented.

The calculations were made under some simplifying assumptions and can be generalized without difficulty to account for additional factors, as e.g. the relative and absolute altitudes above sea level of observers and lights (here assumed to be equal), the change of the luminous intensity of the wind farm lamps with the emission angle, or the spectral dependence of the conversions between Johnson-Cousin V magnitudes and luminances. Also, we have assumed photopically adapted observers (that is, observers who are looking at the lights from windows of lit spaces indoors, or shortly after leaving illuminated areas outdoors). The calculations for mesopic or scotopic adaptation states can be easily done if some additional information about the wind farm light spectra is available, see e.g. [7,23].

In Section 3 it has been implicitly assumed that, although each wind turbine light is seen as an unresolved object, different lights are enough separated angularly among themselves as to ensure that the corresponding PSFs do not significantly overlap in the observer's retina. The objects can then be perceived independently, the attention may shift from one to another [39], the detection thresholds remain unaltered [27], and their magnitudes $m_v$ are correctly described by Eq. (15) using the $I_v$ of each individual lamp. However, on some occasions several distant lights may happen to be angularly very close, as seen from the observer, such that their PSFs substantially overlap in the retina. In that case they would appear as a single and brighter (but still unresolved) object, and its resulting magnitude $m_v$ shoud be calculated by adding first the iluminances $E_v(r)$ produced by each lamp on the eye pupil, Eq. (6), taking into account their possibly different distances and luminous intensities, and then applying Eq. (15) to the total $E_v(r)$ resulting from this sum.

As shown in Section 3, the distance range of the disruptive effects of wind-farm lights is different if expressed in terms of the ground illumination (horizontal illuminance) or in terms of visual perception of the lights (retinal image brightness). Whereas the horizontal illuminance is comparable to or larger than the one produced by the natural moonless starry sky up to distances of a few hundred meter (580 m for 2000 cd lamps), the visual brightness of the lamps competes with the dimmest visible stars of the night sky up to several tens of km (38 km, for the same type of lamps). This difference in range is not unexpected. It is mainly due to the focusing ability of the eye, which concentrates the flux from the pupil illuminance onto a small region of the retina (equal to the eye PSF, for an unresolved light source), proportionally increasing the illuminance on the retinal photoreceptors and hence their photon catches. The retinal images of the lights, then, act as conspicuous visual stimuli even at long distances, and may have attracting or disorienting effects in species that navigate based on cues provided by localized light sources. The horizontal illuminance, in turn, is instrumental for the visual perception of the ground, and the threats and opportunities present on it.

The basic model presented in this work allows for an easy quantification of the visual impact of individual wind farm lights on the nocturnal landscape, by direct comparison of their brightness with the brightness of conspicuous objects of the pristine starry sky. This visibility model is based on well-known first principles, and the results presented here correspond to average atmospheric conditions. The application to observations in field conditions is under way. The main metrological challenge lies in the accurate determination of the aerosol content of the atmosphere at the moment of observation. Whereas daytime data on aerosols are widely available, reliable nighttime data are considerably more scarce.

An issue that has not been addressed here and that deserves further consideration is the quantification of the visual aggregated impact of high numbers of wind farm lights shining simultaneously across large stretches of the horizon, a not uncommon situation in areas with very high density of turbines. Metrics developed for quantifying aggregated visual impacts during daytime [31,36] offer some interesting starting points. Besides its direct application to landscape assessment, it may be of potential interest for professional astronomy, whose ground-based observatories are subjected to an increasing stress by the presence and effects of artificial lights [21,26].

This paper dealt with the quantitative evaluation of the visual effects of wind farm lights, without delving into a detailed discussion of the remediation measures that could be adopted. This discussion requires a comprehensive, broader social approach that deserves a specific work. Classical technological adaptations may not be sufficient nor, in many cases, applicable. Unlike public streetlamps, which are designed to illuminate public spaces but do not require to be seen themselves, the wind-farm light beacons are purposely installed to be seen from long distances in a broad range of directions. This precludes in principle the use of lamp shielding approaches that have been shown to be very useful for reducing the ecological impact of street luminaires [18].

## 5. Conclusions

Wind farm lights are a source of light pollution in the nocturnal landscape. In this work we quantify the impact of individual light sources by comparing their perceived brightness with the brightness of the stars and other conspicuous bodies of the starry night sky. For typical parameters of the lights and the atmosphere our model shows that medium-intensity turbine lights can be brighter than Venus up to ~4 km distance, brighter than α CMa (the brightest star on the nighttime sky) up to about ~10 km, and reach the standard stellar visibility limit for the unaided eye ($m_v = +6.00$) at ~38 km. These results suggest that the visual range of wind farms at nighttime may be considerably larger than at daytime. This factor should be taken into account in environmental impact assessments.

## Acknowledgments and AI disclaimer

No AI tools have been used in this work.